\documentclass[12pt]{article}
\usepackage{graphicx}
\advance\textheight by \topskip
\begin{document}
\tolerance=100000
\thispagestyle{empty}
\setcounter{page}{0}
\def\mpT{p_T \hspace{-1em}/\;\:}
\def\lsim    {\:\raisebox{-0.5ex}{$\stackrel{\textstyle<}{\sim}$}\:}
\def\gsim    {\:\raisebox{-0.5ex}{$\stackrel{\textstyle>}{\sim}$}\:}
\def\mev     {\: \rm MeV}
\def\gev     {\: \rm GeV}
\def\tot     {\: \rm tot}
\def\pp      {$ \rm pp$}
\def\sigtot  {\mbox{$\sigma_{\rm tot}^{\rm pp/ p \bar p}$}}
\def\sigtotpp  {\mbox{$\sigma_{\rm tot}^{\rm pp}$}}
\def\rs{\mbox{$\sqrt{s}$}}
\def\pbarp{\mbox{$\rm  \bar{p}p$}}
\def\ptmin{\mbox{$p_{tmin}$}}
\newcommand{\ra}{\rightarrow}
\newcommand{\ba}{\begin{array}}
\newcommand{\ea}{\end{array}}
\newcommand{\beqa}{\begin{eqnarray}}
\newcommand{\eeqa}{\end{eqnarray}}
\newcommand{\be}{\begin{equation}}
\newcommand{\ee}{\end{equation}}

\newcommand{\comment}[1]{}

\vspace*{\fill}
\begin{flushright}
{IISc-CHEP/09/07}\\
\end{flushright}
\begin{center}
{\Large \bf Total cross-section and rapidity gap survival probability
at the LHC through an eikonal with soft gluon resummation} \\[0.25cm]
\end{center}
\begin{center}
{\large
Andrea  Achilli$^{1}$, Rohit Hegde$^{2}$,
Rohini M. Godbole$^{3}$, Agnes Grau$^{4}$,
        Giulia  Pancheri$^{1}$ and Yogi Srivastava$^{5}$
 }\\[0.25 cm]
\end{center}
\begin{center}
$^1$INFN, LNF, P.O. Box 13, I-00044 Frascati, Italy.
         \\[0.2cm]
$^2$ Department of Physics, University of Texas at Austin, 1 University Station
 C1600, Austin, Tx 78712-0264\\[0.2cm]
$^3$ Centre for High Energy Physics, Indian Institute of Science,Bangalore,
560012,India.  \\[0.2cm]
$^4$Departamento de Fisica Teorica y del Cosmos, Universidad de Granada, Spain.
         \\[0.2cm]
$^5$Physics Department and INFN, University of Perugia,
             Perugia, Italy.
\end{center}
\vspace{0.4cm}
\abstract{
New results are presented for total ${\rm pp}/ {\rm \bar p p}$ cross-sections, 
in the framework of our QCD based model (GGPS). This is an improved eikonal 
mini-jet model, where soft gluon radiation tames the fast energy rise normally 
present in mini-jet models.  We discuss the variability in our predictions
and provide a handy parametrization of our results for the LHC. We find that
our model predictions 
 span the range $\sigma_{tot}^{LHC}=100 ^{+10}_{-13}\ mb$.
While this matches nicely with the range of most other models,
it does not agree with  recent  ones which include a "hard" Pomeron, even 
though our  model does include hard scattering. We compute the survival 
probability for Large Rapidity Gap (LRG) events  at the LHC and at the
Tevatron. These events are relevant, for example, for Higgs signal in the $WW$ 
fusion process.  We also explore whether measurements of the  total 
cross-sections at the LHC can help us sharpen the model parameters and 
hence estimates for  these  survival probabilities, further. }

\section{Introduction}
In this letter we discuss the upcoming measurements of the total proton-proton
cross-section at LHC, in the context of a QCD based model, which may be
used to shed light on the role played by soft gluon resummation
in the infrared limit \cite{Corsetti:1996wg}. 
We work in an Eikonalised mini-jet model and  achieve unitarisation 
through an Eikonal, where the energy dependent impact parameter distribution 
is calculated in a QCD based model using realistic parton densities.
This model, for specific values of the parameters, chosen by confronting
it with available data, gave a value $\sigma_{tot}^{LHC}=100.2 \  mb$ 
\cite{Godbole:2004kx}.  
It is our purpose 
here to present  cross-section estimates at LHC  for a full range of parameter 
choices and   provide a useful parametrisation of these 
for comparison  with the LHC data. It can then be further used to describe 
other soft quantities such as the  underlying event distributions and rapidity 
gap survival probabilities.  Note  that 
a reliable prediction of total non-diffractive cross-section is essential
for a correct projection of the expected underlying activity at the LHC, 
which in turn is required at times to ensure the correct 
extraction of new physics from the LHC data. Surely we
will have to depend -at the initial stages of LHC- upon predictions
based on our current understanding of these matters. 
There exists a close relationship between the energy dependence of
the total cross-section and the size as well as the
energy dependence of the survival probability of the large rapidity
gaps (LRG).  These are regions in angular phase space devoid of any
particles which might exist in events in  \pp\ /\pbarp\ reactions,
where a colour singlet state is produced and there exist no color connections
between the two colliding hadrons ~\cite{Dokshitzer:1987nc,Dokshitzer:1991he,
Bjorken:1991xr,Bjorken:1992er}.
Such events are predicted, for example, when a Higgs boson is produced 
through $W W$ fusion.  This unique signature can be used very effectively, 
also in searches of other color singlet states which are sometimes predicted in
various Beyond the SM (BSM) scenarios and  which can also  give rise to
events with large rapidity gaps.  Both the events with LRG and the survival 
probabilities 
of the rapidity gaps continue to be a  subject of intense study for this reason
~\cite{Dokshitzer:1987nc,
Dokshitzer:1991he,Bjorken:1991xr,Bjorken:1992er,
Chehime:1992bp,Fletcher:1993ij,Khoze:1997dr,
Gotsman:1998mm,Khoze:2000cy,Block:2001ru,Gotsman:2005wa,
Luna:2006qp}.

The plan of the letter is as follows :  in Section 2
we present the salient features of the above model, and
explore  the dependence of results  for total cross-section
on the QCD inputs such as the parton density functions (PDFs).
In section 3 we explore the various perturbative, nonperturbative
inputs and parameters of the model and  compare the predictions of 
our model with the data. We use the results of this exploration of model 
parameters
to obtain an estimate of the ``theoretical'' uncertainty.
We also provide a simple parametrisation  of our model and a table with 
numerical estimates for the LHC.
In Section 4  we present our expression for the survival probability for 
LRG and evaluate it within our model, using a representative set of 
parameters, and comparing it with predictions from other approaches.

\section{Eikonal Mini-jet Model  model with soft gluon resummation in impact parameter space.}
In this section, we recall briefly some of the relevant details of our
model~\cite{Corsetti:1996wg,Godbole:2004kx}.
We shall then apply it 
to estimate the total cross-section at LHC and, in the last section,
to predict Survival Probabilities for Large Rapidity Gaps (SPLRG) at LHC.

Generically, the total
cross-section can be written as
\begin{equation} \label{DL4}
\sigma_{tot}^{AB}(s)= \sigma_{soft}(s) + \vartheta (s - \bar{s})
\sigma_{hard}(s)
\end{equation}
with $\sigma_{soft}$ containing a constant ( the ``old'' Pomeron
with $\alpha_P(0)\ =\ 1$) term plus a (Regge) term decreasing as
$1/\sqrt{s}$,  with an estimate for the constant $\sim 40$
mb~\cite{Godbole:2004kx}. In the mini-jet model
the rising part \cite{cline}
of the cross-section $\sigma_{hard}$ is provided by
jets which are calculable by perturbative QCD
\cite{rubbia,gaisser,pancheri}, obviating (at least in
principle) the need of an arbitrary parameter $\epsilon$
which controls the rise as in \cite{Donnachie:1992ny,Landshoff:2005rg}. The
increase in \sigtot\ with energy is driven by the rise with energy
of $\sigma_{jet}^{AB}$ which is given by \be \label{sigjet}
\sigma^{AB}_{\rm jet} (s;p_{tmin}) = \int_{p_{tmin}}^{\rs/2} d p_t \int_{4
p_t^2/s}^1 d x_1 \int_{4 p_t^2/(x_1 s)}^1 d x_2 \sum_{i,j,k,l}
f_{i|A}(x_1,p_t^2) f_{j|B}(x_2, p_t^2)~~
 \frac { d \hat{\sigma}_{ij}^{ kl}(\hat{s})} {d p_t}.
\ee Here  subscripts $A$ and $B$ denote the colliding particles 
($p$ and/or  $\bar p$), $i, \ j, \ k, \ l$ the partons and $x_1,x_2$ the 
fractions of the parent particle momentum carried by the parton. 
$\sqrt{\hat{s}} = \sqrt{x_1 x_2 s}$  and $\hat{ \sigma}$ are the
centre of mass energy of the two parton system and the hard parton 
scattering cross--section respectively.
Let us note that parton density functions (PDF's) in the proton, extracted from
QCD analysis of a variety of data and the basic elements of perturbative QCD 
such as the elementary subprocess cross-sections, are the only inputs
needed for the calculation of $\sigma_{\rm jet}^{\rm AB}$. 
Needless to say one uses the DGLAP evoluted, $Q^2$ dependent PDF's.
The rate of rise with energy of this cross-section is determined by
$p_{tmin}$ and the low-x behaviour of the parton densities.
As often discussed, this rise  is much steeper
than can be tolerated by the Froissart bound. Hence the mini-jet cross-section
is imbedded in an eikonal formulation \cite{durand}, namely
\begin{equation}
\sigma^{AB}_{tot}=2\int d^2{\vec b}[1-e^{-\Im m\chi^{AB}(b,s)}
cos (\Re e\chi^{AB}(b,s))]
\label{stot}
\end{equation}
where $ 2\Im m~\chi^{AB}(b,s)=n^{AB}(b,s)$ is the average number of
 multiple collisions assumed to be
Poisson distributed \cite{daniele}.
The quantity $n(b,s)$  has contributions coming from both soft and
hard physics and we write it as
\begin{equation}
n^{AB}(b,s)=n^{AB}_{soft} (b,s) + n^{AB}_{hard}(b,s)
\label{nsplit}
\end{equation}
By construction,  $n_{hard}$  includes only  parton-parton  collisions where
the scattered partons have $p_t\ge p_{tmin}$, the cut off in the mini-jet
cross-section. It follows that all other collisions
are included in $n_{soft}$.

In the standard formulation of the eikonal
for the total cross-section, $n(b,s)$ is assumed to
factorize into a $b$-dependent overlap
function $A(b)$,  which is a measure of the overlap in the transverse
plane of the partons in the colliding beams, and an $s$--dependent
(soft +  jet) cross-section.
 Most eikonal models,
including  QCD driven ones of  {\it Refs.}  \cite{Block:2001ru,Luna:2006qp,bghp}, propose
 a functional $b$-dependence derived from the
Fourier transform of the electromagnetic Form Factor (FF).
However, as already noted some time ago in ~\cite{durand},
these eikonal models would lead to  too steep a rise of the total
cross-section with energy, if one would use actual QCD
mini-jet cross-sections with  a fixed $p_{tmin}$.  This was shown in detail
for GRV densities \cite{Gluck:1991ng} in \cite{Godbole:2004kx}.

An altogether different approach is to relax  the $b-s$ factorization and 
assume that the $b$-distribution, $A(b)$, is energy
dependent. This is physically what one expects, since when  two hadrons
collide the matter distribution cannot stay constant, rather the partons
influence each other's path. We  make this idea of the shift in the path of 
the  parton more quantitative by 
modeling  the $b$-distribution as the Fourier transform of the change in 
collinearity of the partons  due to soft gluon emission before the collision.
Let $A(b)=A_0$ at time $t=-\infty$,
before the collision. At this time the partons do not
yet influence each other and
stay in some ideal ''hadronic'' configuration. This configuration gets modified
as they approach each other and soft gluon emission  takes
place as  partons feel each other's color field and
scatter.
Let $\Pi({\bf K}_t)d^2{\bf K}_t$ be the probability distribution that a pair of partons acquires a transverse 
acollinearity ${\bf K}_t$ because of soft gluon emission before the collision.
Then the change in the static ideal $b$-distribution $A_0$ in our model
is calculated as the Fourier transform of this 
probability  and
the quantity $A_0$
is fixed by the normalization requirement,
namely that the probability of finding two partons at a distance
{\bf b}  from each other must be 1 when we sum over all possible values.
This gives
\be A(b,s)=A_0 \int d^2 {\bf K}_t e^{-i{\bf K}_t\cdot {\bf b}}\Pi({\bf K}_t)=\frac { e^{-h(b,q_{max})}}
{\int d^2{\bf b}  e^{-h(b,q_{max})}}\equiv A_{BN} (b,q_{max})
\label{adb}
\ee
where the function  $h(b,q_{max}) $
is obtained from soft gluon resummation techniques \cite{Corsetti:1996wg}.
Because, in general,  soft gluon emission is   energy dependent,
the assumption of the factorization into a $b$--dependent
piece and an $s$--dependent piece is automatically relaxed.
 We denote the corresponding overlap functions by
 $A_{BN} (b,q_{max})$, where  BN stands for Bloch-Nordsiek,
to remind us of the origin of the soft guon resummation factor and notice
that it depends  (i) on the energy, (ii) the kinematics of the subprocess and
(iii) the parton densities.
Depending on how one models $q_{max}$,  the
rapid rise in the hard, perturbative jet part of the
eikonal can  then be tamed, into the experimentally observed mild increase, by
soft gluon radiation whose maximum energy ($q_{max}$) rises slowly
with energy.

Before evaluating $n_{hard}$, we point out  that  the evaluation of
$A_{BN}$  through the function $h(b,s)$ in Equation \ref{adb}, involves
$\alpha_s$ in the infrared region \cite{Corsetti:1996wg, Godbole:2004kx}. 
In our model, an important role is played by the integral of $\alpha_s$ 
down to zero momentum gluons.  This is a non-perturbative region and we  
model the  behaviour as a power law, 
namely $\alpha_s(k_t) \approx  k_t^{-p}$ as $k_t \to 0$.
As noticed before \cite{dokhalphas},  what is observable are only 
moments of $\alpha_s$ and not the single vertex. In order to have a 
continuous analytic expression valid also in the perturbative region, 
we then 
 use a phenomenological
form inspired by the Richardson potential \cite{Corsetti:1996wg}, namely
\begin{equation}
\alpha_s(k_t^2)={{12 \pi}\over{33-2N_f}} {{p}
\over{\ln[1+p({{k_t}\over{\Lambda}})^{2p}]}}
\end{equation}
This parametrization of the infrared behaviour of $\alpha_s$ involves the
parameter $p$ which for the Richardson potential is $1$, but which we take 
always as $less\  than\  1$ for the integral to be convergent.

One can now use $q_{max}$ values (obtained from kinematical 
considerations \cite{greco,kazimierz}) to calculate the impact parameter 
distribution for the hard part of the eikonal, 
namely $A_{BN}(b,q_{max}^{jet})$,
and then $n_{hard}(b,s)$, using the mini-jet cross-sections. Notice that for a 
given set of QCD parton densities, one obtains corresponding values for 
$\sigma_{jet}$ and $q_{max}^{jet}$. 
The interplay of these quantities and their dependence 
upon the densities and  $p_{tmin}$ has been explicitely discussed in 
\cite{kazimierz}.
The ''hard'' part is thus fully determined.
 The
 subsequent step of obtaining the full $n(b,s)$ and
its eikonalization requires
finding appropriate values for the soft part of the
eikonal.
  The 'soft' part, determined by
non-perturbative dynamics, is  modeled as follows: $n_{soft}$ is
factorized into a non-rising
soft cross-section $\sigma_{soft}$ and
$A_{soft} = A_{BN}(b,q_{max}^{soft})$.
The non-perturbative, soft part of the eikonal
includes only limited low energy gluon emission and leads to the
initial decrease in the proton-proton cross- section.
$q_{max}$ is assumed to be
the same for the hard and soft processes at low energy ($\sim 5\ GeV$),
parting
company around $10$ GeV where hard processes become important.

 Thus, neglecting the real part of the eikonal, one can now calculate
the total $pp$ and $p{\bar p}$ cross-section with Equation \ref{stot}
and  $n(b,s)$ as given
below:
\begin{equation}
n(b,s)  = A_{BN}(b,q_{max}^{soft}) \sigma_{soft}^{pp,{\bar p}}+
A_{BN}(b,q_{max}^{jet})
 \sigma_{jet}(s;p_{tmin}),
\end{equation}
where
\begin{equation}
\sigma_{soft}^{pp}=\sigma_0, \ \ \ \ \ \ \ \ \ \ \
\
\sigma_{soft}^{p{\bar p}}=\sigma_0
(1+{{2}\over{\sqrt{s}}})
\end{equation}

The three parameters of the model so far are $p_{tmin}$,
$\sigma_0$ and p. Values of
$p_{tmin}, \sigma_0$ and $p$  which give a good fit to the data with
the GRV parametrisation of the proton densities~\cite{Gluck:1991ng}
are $1.15$ GeV,  $48$ mb  and $3/4$ respectively, as presented in
Ref.~\cite{Godbole:2004kx}. These values are  consistent with the
expectations
from a general argument~\cite{Godbole:2004kx}.
\begin{figure}
\begin{center}
\includegraphics[scale=0.40]{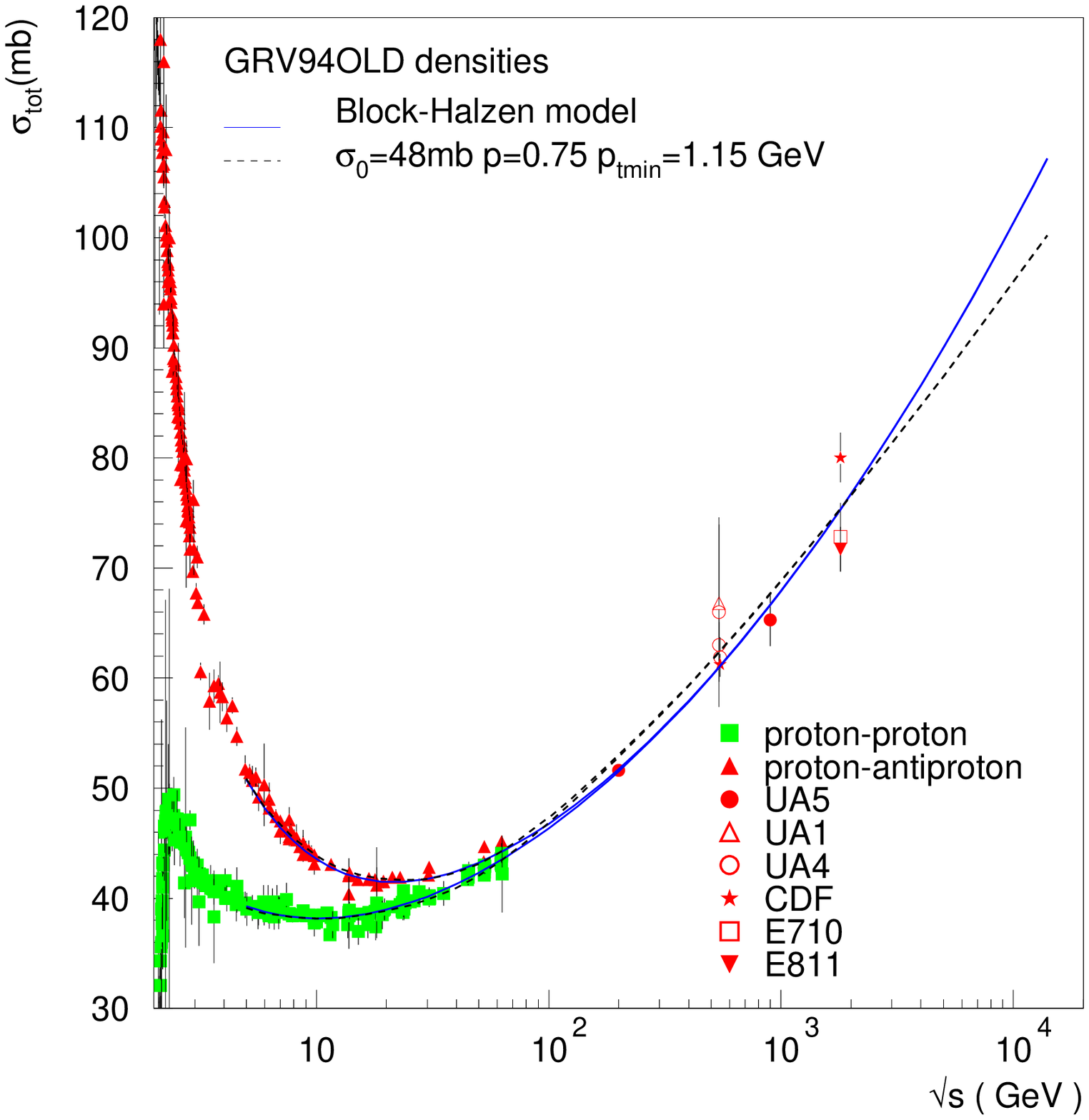}
 \hspace{.1cm}
\includegraphics[scale=0.40]{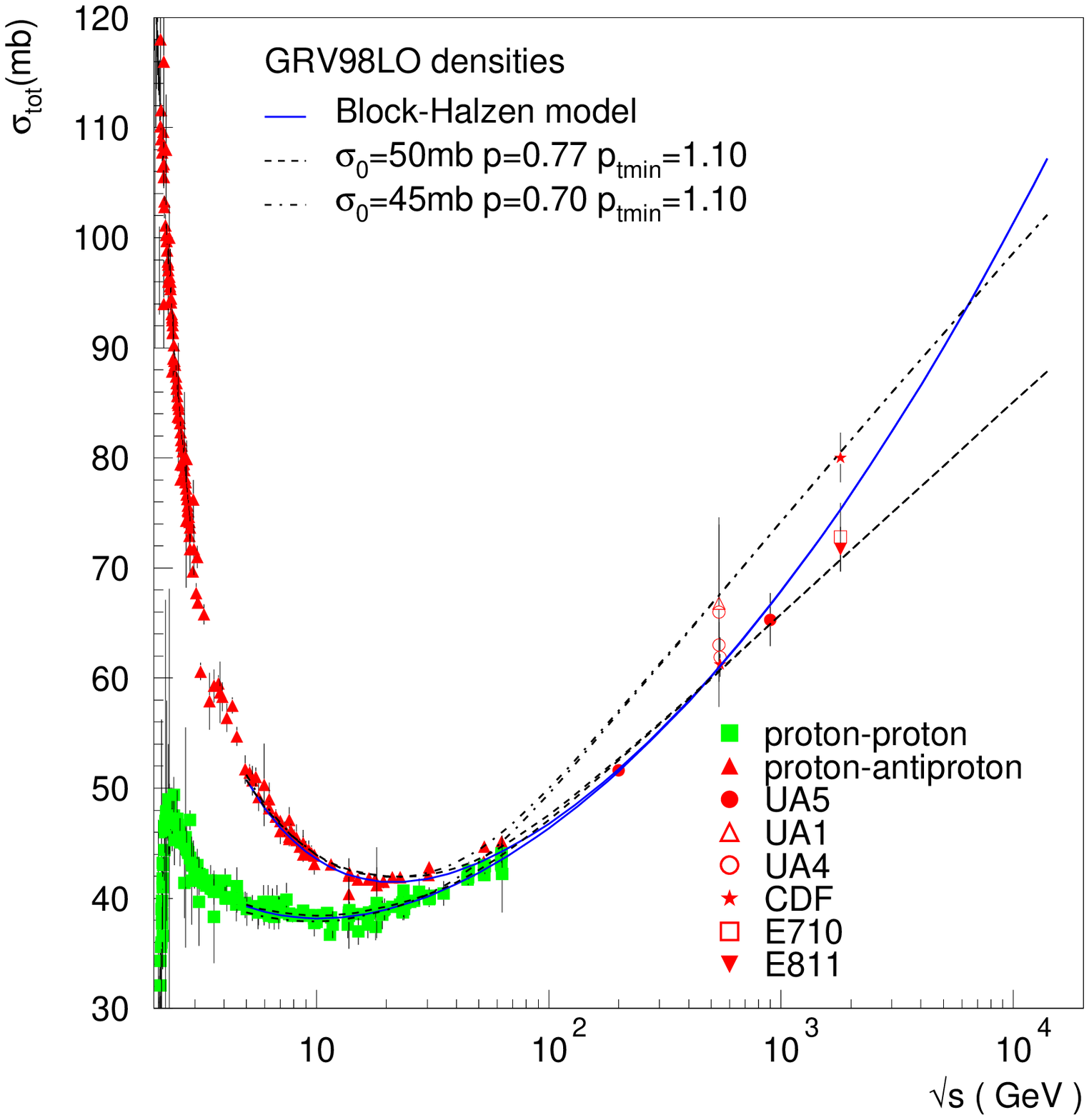}
 \hspace {.1cm}
\caption{Comparison of the GGPS predictions for GRV and GRV98lo
densities with data and with the BH \protect\cite{Block:2005ka}
prediction. The parameter set used for the GGPS model are also
shown. }
\label{bgrv94old}
\end{center}
\end{figure}
Figure ~\ref{bgrv94old} compares the predictions of GGPS with data
\cite{PDG,UA1,UA4,UA5,Amos:1991bp,Avila:1998ej,Abe:1993xy}  as well
as the one obtained in \cite{Block:2005ka} 
by phenomenological  considerations along with unitarity
and factorisation.  It should be noted here that,
in contrast to other models which employ
the eikonal picture, in our model the eikonal is
determined in terms of just these three parameters along with  the
parton densities in the proton and the QCD subprocess
cross-sections. We expect these {\it favourite } values to change
somewhat with the choice of parton density functions.  Since we
are ultimately interested in the predictions of the model at TeV
energies, we need PDF parametrisation which cover both the small
and large $Q^2$ range and are
reliable up to rather small values of $x (\sim 10^{-5})$. Further, since our
calculation here is only LO, for consistency we have to use LO densities.

We notice \cite{kazimierz} that not all sets of PDF's
return the correct energy dependence for the total cross-section.
This is clearly due to the fact that our model
probes down to very low-x values for the mini-jet cross-sections.
As the energy increases, the fixed value of $p_{tmin}$ amounts to
receiving  contributions from $x_{gluon}\approx 10^{-5}$
and not all densities have the same behaviour at such low x-values. 
Recall that only limited amount of data are available in the small--$x$
region and for $x_{gluon} \lsim 10^{-5}$, almost all the PDF's 
use extrapolation of the parton densities at higher values of $x$ 
where they are obtained by fits to the data.
In particular, we note that 
the CTEQ \cite{CTEQ} densities lead to total cross-sections which 
start decreasing beyond the Tevatron energy range, thus differing 
from all the other densities. Within our model, $q_{max}^{CTEQ}$  is seen 
to rise to higher values, the consequent decrease in the cross-section
more than compensates for the rise due to the minijets, thus leading to 
cross-sections decreasing with energy.
This shows the interplay between the densities and soft  gluon emission.
On the other hand, it is comforting to see that other commonly used densities,
such as GRV and MRST  \cite{Gluck:1994uf,Gluck:1998xa,Martin:2002dr}, give 
results in the same range and with acceptable energy behaviour up to cosmic 
ray energies. This characterizes our model as being stable versus most 
available density types. From now on we shall only employ the GRV and MRST
densities in further analysis.

\section{Predictions for \sigtot\ at LHC}
Having described the role played by PDF's in a computation of total 
cross-sections at LHC, we  explore for a range of
PDF's~\cite{Gluck:1994uf,Gluck:1998xa,Martin:2002dr}
different inputs for $p$, $p_{tmin}$ and $\sigma_0$.
For each PDF, the onset of the rise fixes $p_{tmin}$, $\sigma_0$
controls the normalization and $p$
determines the slope of the rising part of the cross-section (as well as
the normalization through $A_{BN}$). We
find that it is possible to get a satisfactory description of all
current data, for all choices of PDF's considered, namely MRST and GRV.
 The
corresponding range of values of $p_{tmin}, \sigma_0$ and $p$ are
given in Table 1, 
together with expected values of $\sigma_{tot}$  
for the LHC as well as the expectations for  $<|S^2|>$, the probability of 
survival for large rapidity gaps. The latter will be discussed in the next 
section.

\begin{table}
\begin{footnotesize}
\begin{center}
\begin{tabular}{|c|c|c|c||c|c|}
\hline
&&&&&\\
PDF&$p_{tmin}$ (GeV) & $\sigma_0$ (mb)&p&$\sigma_{tot}^{LHC}$ (mb)& $<|S^2|>$\\
\hline
&&&&&\\
GRV \cite{Gluck:1991ng}&1.15&48&0.75& 100.2&0.101 \\
\hline
&&&&&\\
GRV94lo\cite{Gluck:1994uf} &1.10&46&0.72&103.82&0.127  \\
&1.10&51&0.78& 89.65&0.089\\
\hline
&&&&&\\
GRV98lo\cite{Gluck:1998xa} &1.10&45&0.70& 102.05&0.154        \\
&1.10&50&0.77& 87.83&0.106\\
\hline
&&&&&\\
MRST\cite{Martin:2002dr} &1.25&47.5&0.74& 95.92&0.123\\
&1.25&44&0.66& 110.51 &0.172   \\
\hline
\end{tabular}
\caption{
Values of \protect{$\sigma_{tot}$} for
 $p_{tmin}$, $\sigma_0$ and $p$ corresponding to different
parton densities in the proton, for  which GGPS
 gives a satisfactory description of
\sigtot\ .
}
\end{center}
\end{footnotesize}
\end{table}

We now  compare the expectations from different models. The DL
parameterisation~\cite{Donnachie:1992ny}
\begin{equation}
\sigma_{tot}(s)=Xs^\epsilon+Ys^{-\eta},
\label{DL}
\end{equation}
is a fit to the existing data with $\epsilon =0.0808, \eta
= 0.4525$. This fit has been extended to include a 'hard'
pomeron~\cite{DLnew} due to the discrepancy between different data
sets. The BH model~\cite{Block:2005ka} gives a fit to the data using
duality constraints. The BH fit for $\sigma^{\pm} = \sigma^{\bar p
p}/ \sigma^{p p}$ as a function of beam energy $\nu$, is given as
$$
\sigma^{\pm} = c_0 +c_1 \ln(\nu/m) + c_2 ln^2 (\nu /m)
+ \beta_{P'} (\nu/m)^{\mu-1} \pm \delta (\nu/m)^{\alpha -1},
$$
where  $\mu = 0.5, \alpha = 0.453\pm 0.0097$ and all the other
parameters (in mb) are $c_0 = 36.95, c_1 = -1.350 \pm 0.152, c_2 =
0.2782 \pm 0.105, \beta_{P'} = 37.17, \delta = -24.42 \pm 0.96$
from \cite{Block07}. The fit obtained by Igi et al.
\cite{Igi:2005jm}, using FESR \cite{fesr}, gives LHC predictions
very similar to those given by the BH fit. Predictions have been advanced 
by Luna and Menon \cite{Luna07} using fits to low energy proton-proton  
and cosmic ray data  from Akeno \cite{Honda:1992kv} and Flye's Eyes
\cite{Baltrusaitis:1984ka}. Avila, Luna and Menon  give also
fits ~\cite{Avila:2002tk} using analyticity arguments and different sets of 
cosmic ray data. Depending on the analytic expression used and set or model 
used to extract the cross-sections from the cosmic ray data, their 
cross-section predictions at LHC vary. Finally, using an eikonal model 
inspired  by
BGHP \cite{bghp}, Luna and collaborators \cite{luna05}  
predict $\sigma_{tot}^{LHC} =102.9\pm 7.1\ mb$. In the framework of
the COMPETE program, Cudell et al~\cite{Cudell:2002xe} give
predictions for the LHC energies by extrapolating fits obtained to
the current data based on an extensive study of possible analytic
parametrisations, using again constraints from unitarity,
analyticity,
factorization, coupled with a requirement that the cross-section
asymptotically goes to (i) a constant, (ii) as $\ln s$ or as (iii)
$\ln^ 2 s$. Their central value of the fit has no $\ln{s}$ term, and it 
predicts $\sigma^{pp}(LHC)=111.5\pm 1.2^{+4.1}_{-2.1} (mb)$, where the 
systematic errors come from the choice in the fit between CDF and E710/E811 
data at the Tevatron.  Recently, Cudell and Selyugin
\cite{cudsel} have considered predictions from a model with both a
hard and a soft Pomeron term, leading to a cross-section of the order of
140 mb at LHC energies.

\begin{figure}
  \begin{center}
    \includegraphics[scale=0.50]{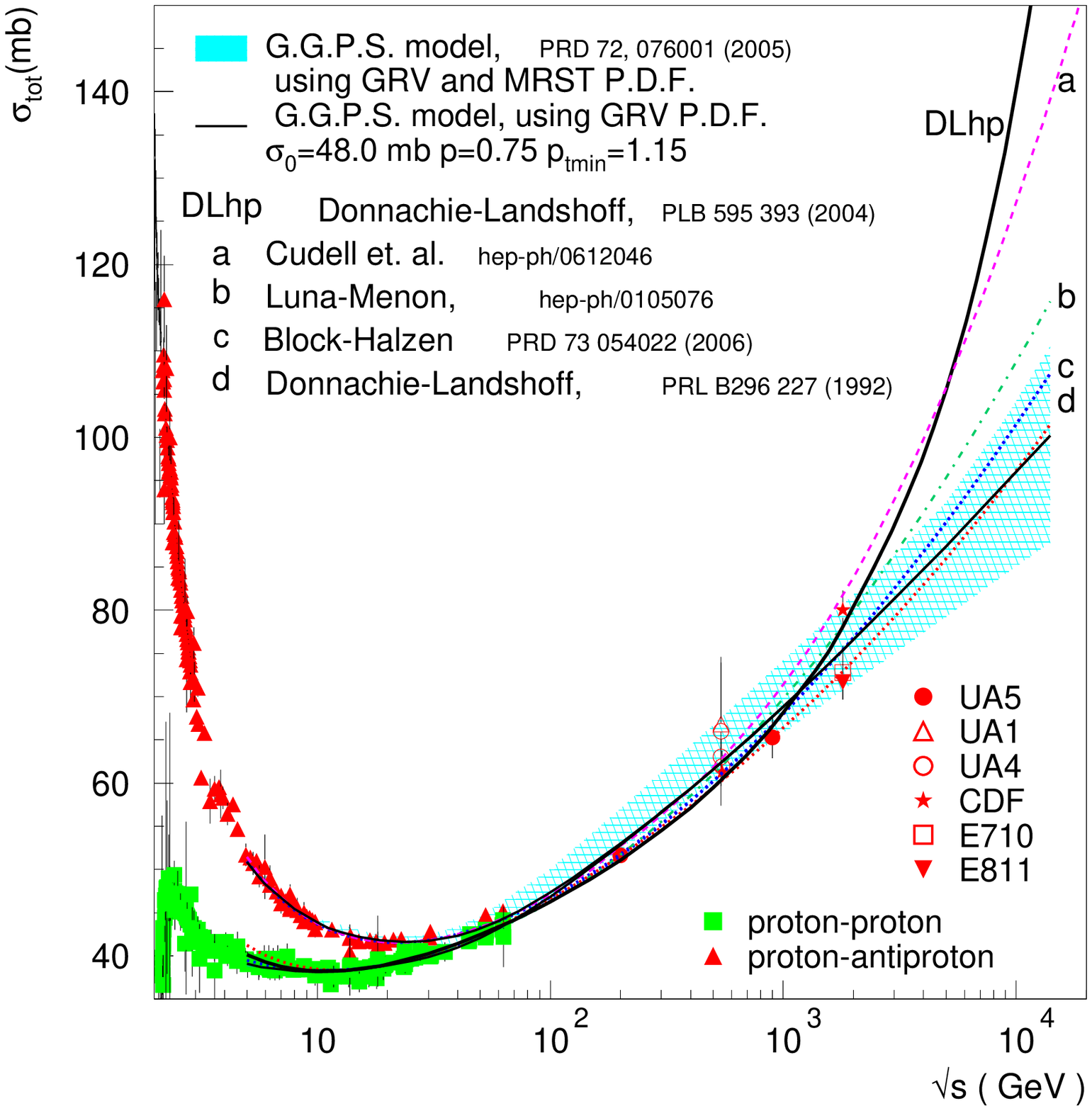}
\caption{Predictions for \protect\sigtot\ in various models. The
shaded area gives the range of results from GGPS~\protect\cite{Godbole:2004kx}
the solid line giving the prediction obtained using the GRV
parton densities~\protect\cite{Gluck:1991ng} in the model.
\comment{
The
long-dashed dotted curve ($d$), indicates the predictions for the DL
fit~\protect\cite{Donnachie:1992ny}.  The dotted(BH) curve ($c$) and
the uppermost dashed curve ($a$), are the results of two analytical
models incorporating constraints from unitarity and analyticity,
from~\protect\cite{Block:2005ka} and ~\protect\cite{Avila:2002tk},
respectively. The prediction obtained by Igi et al, using FESR
follows very closely that given by the BH curve. Further the short
dash dotted curve ($b$), is the result of a fit by the COMPETE
collaboration~\protect\cite{Cudell:2002xe}.}
} \label{csection}
\end{center}
\end{figure}

Figure~\ref{csection} summarizes the predictions of some of the  models
described in the previous paragraphs.  Curve (d),
indicates the predictions of the standard Regge-Pomeron
fit~\protect\cite{Donnachie:1992ny}, while the new fit with a hard
Pomeron term  is labeled (DLhp). The two curves labeled (c) and (b) are the results of  fits with the analytical
models 
from~\protect\cite{Block07}
and ~\protect\cite{Luna07} respectively.
 The short
dash dotted curve (a) is from \cite{cudsel} with a
hard Pomeron term.
The shaded area gives the range
of predictions in the GGPS model  with soft gluon
resummation~\protect\cite{Godbole:2004kx}, the different PDF's used
giving the range as described  earlier, and the solid line at the center of
the band being  the one obtained with the GRV parton
densities~\protect\cite{Gluck:1991ng} and other parameters as 
in \cite{Godbole:2004kx}.  We  see that the range of results from GGPS
for  LHC spans other predictions based on models using unitarity, 
factorization, analyticity and  fits to the current data. 
The  
predictions shown fall in two groups, those with an explicit "hard" Pomeron 
and those based on analyticity and unitarity constraints which seem 
consistent with each other.  We are in disagreement with models which 
incorporate a "hard" Pomeron. Indeed our model has a hard QCD component, the 
mini-jet cross-section discussed in the previous section, but soft QCD 
emission tempers it and brings the fast rise back to a smooth behaviour.
In the end it predicts a growth with energy not faster than $\ln^2{s}$,  
as we shall see in the following.  

\begin{table}
\label{parampp}
\begin{footnotesize}
\begin{center}
\begin{tabular}{|c|c|c|c|c|c|c|c|c|}
\hline
\multicolumn{4}{|c|}{}              &          &           &  &           &
\\
\multicolumn{4}{|c|}{proton-antiproton} &$a_0$ (mb)& $a_1$ (mb)&b & $a_2$ (mb)&
$a_3$ (mb)\\
\multicolumn{4}{|c|}{}&&&&&\\
\hline
&&&&&&&&\\
$\sigma_0$ (mb)&$p_{tmin}$ (GeV)&p&&&&&&\\
&&&&&&&&\\
\hline
&&&&&&&&\\
50&1.10&0.77 &Lower edge& $17 \pm 1$  & $122 \pm 6$ & $-0.5$ & $2.7 \pm 0.1$ & $0.054 \pm 0.036$ \\
\hline
&&&&&&&&\\
44&1.25&0.66&
Top edge& $19 \pm 2$ & $127 \pm 4$ & $-0.5$ & $1.9 \pm 0.3$ & $0.149 \pm 0.013$  \\
\hline
&&&&&&&&\\
48&1.15&0.75&
Center& $20 \pm 1$ & $125 \pm 6$ & $-0.5$ & $1.6 \pm 0.1$ & $0.135 \pm 0.012$ \\
\hline \hline \multicolumn{4}{|c|}{}              &          &
&  &           &\\
\multicolumn{4}{|c|}{proton-proton} &$a_0$ (mb)& $a_1$ (mb)&b & $a_2$ (mb)&
$a_3$ (mb)\\
\multicolumn{4}{|c|}{}&&&&&\\
\hline
&&&&&&&&\\
$\sigma_0$ (mb)&$p_{tmin}$ (GeV)&p&&&&&&\\
&&&&&&&&\\
\hline
&&&&&&&&\\
50&1.10&0.77&Lower edge& $17 \pm 2$  & $65 \pm 7$ & $-0.5$ & $2.7 \pm 0.4$ & $0.054 \pm 0.017$ \\
\hline
&&&&&&&&\\
44&1.25&0.66&Top edge& $19 \pm 2$ & $63 \pm 5$ & $-0.5$ & $1.8 \pm 0.3$ & $0.148 \pm 0.015$ 1\\
\hline
&&&&&&&&\\
48&1.15&0.75&Center& $20 \pm 1$ & $66 \pm 5$ & $-0.5$ & $1.6 \pm 0.2$ & $0.135 \pm 0.013$ \\
\hline
\end{tabular}
\caption{
Values of  $a_0,a_1,a_2,a_3$ and $b$
parton densities in the proton, for  which GGPS
Ref.~\protect\cite{Godbole:2004kx} gives a satisfactory description of
\sigtot\  .  }
\end{center}
\end{footnotesize}
\end{table}

The top edge of the GGPS prediction is obtained for the MRST
parametrization whereas the lower edge for the GRV98lo, with other
parameters as in Table 1. In GGPS, we have parametrised the maximum growth
with a $\ln^2 s$ term.  We find it to give a better representation of our 
results than a term of the Regge-Pomeron type.
We  give fits to our results for \sigtot\  of the form, \be \label{emmfits}
\sigtot = a_0 + a_1 s^{b}  + a_2 \ln (s) + a_3 \ln^2 (s). \ee
In these parametrizations
we have constrained the $\log^2 s$ term to have a positive coefficient, 
whereas in \cite{Pramana} this coefficient had been let free to assume
either sign. We show the corresponding GGPS model parameters
$\sigma_0, p_{tmin}$ and $p$  in Table 2.
The corresponding PDF's used in the calculation of $\sigma_{jet}$ can be read 
from Table 1 for the given set of parameter values.

Once the LHC measurements for the total cross-section will have indicated the
best parameter set to use, the model  can be used to  calculate  the 
$b$-distributions, namely average number of collisions, shape of the overlap 
function, etc. at the LHC~\cite{Buttar:2006zd}.

\section{Large Rapidity Gap Survival Probability}
We now employ our model to estimate the survival probability for LRG.
As mentioned already, events with LRG may arise as a signal of (say) Higgs
bosons produced through WW fusion. The importance of the $WW$ fusion
channel for the production of the Higgs boson at LHC to enhance the
potential of LHC towards discovering a 'light' Higgs and
its properties in detail cannot be
overemphasized ~\cite{higgsreview}. The studies in this channel
crucially use the LRG to increase signal/background ratio.
But as Bjorken~\cite{Bjorken:1992er}  pointed out, it is important to
estimate the probability that ordinary QCD processes, including
bremmstrahlung radiation, and the soft
spectator jet activity will not fill this gap in the angular space
with hadrons. The first part can be computed using perturbation
theory~\cite{Khoze:1997dr,Martin:1997kv,Oderda}, it is the second
part corresponding to the soft rescattering contribution that
requires non-perturbative techniques.

Let  the cross-section for
the hard process $AB\to H$ be calculated through a convolution of the parton
densities in the transverse impact parameter plane as \be
\sigma_{\rm H}(s) =\int d^2{\bf b}\ A^{AB}({ b},s) \sigma_H(b,s)
\ee
where
 $A^{AB} ( b, s)$ is the transverse overlap function for the
two projectiles A and B.

Then the gap survival
probability~\cite{Bjorken:1991xr,Bjorken:1992er} is given by
\be
< |S | ^2 > = \frac{\int d^2 {\bf b}\  A^{AB}( b ,s) | S( b) |^2 \sigma_H(b,s)}
{\int d^2 {\bf b}\  A^{AB}( b,s)\sigma_H(b,s)}.
\ee
Here $| S( b) |^2$ is the probability that the two hadrons A,B go
through
each other without an inelastic interaction
if they arrive at an impact parameter
$\bf b$ and $A^{AB}(b,s)$ is the distribution
in impact parameter space for non-jet
like interactions.

In the eikonal formulation used in \cite{Block:2001ru} and more
recently in \cite{Luna:2006qp},  this probability is given by
$| S(b) |^2\ =\ e^{-2\Im m \chi( b,s)}$.
Since it is precisely this eikonal function that is also involved
in the calculation of the total cross-section \sigtot , one can then
use it to calculate the above mentioned survival probability. In the
hypothesis  that the hard process can be factorized out of
the $b$-integration,
 the
net survival probability then is given by \be <| S | ^2 > = \int d^2
{\bf b} \ A^{AB} ({ b},s)\  e^{-2\Im m \chi( b,s)}. \ee Here, the
transverse overlap functions are always assumed to be normalized to
unity. The impact parameter distribution 
which was used in the QCD inspired model
of ~\cite{Block:2001ru} and \cite{Luna:2006qp},
corresponded to the term for quark scattering,
one of three terms used to parametrize the $b$-distribution of the eikonal.
For us it is different and we shall compare our results with these models,
as well as with other predictions in the literature.

We are looking at the probability of survival of large rapidity gaps which
are present in production of colour singlet state (like the Higgs boson
production via WW fusion) without an accompanying hard QCD process.
Our model has both soft and hard components, with hard parton scattering
cross-section for which $p_t\ge p_{tmin}$.
To exclude hard interactions,
for LRG,
we only need to
consider the $b$-distribution of ``soft'' events, where the
hadronic activity is due to collisions with $p_t\le p_{tmin}$.
Thus, our model automatically selects  processes, with
 very low $p_t$
through the soft $A_{BN}(b,q_{max}^{soft})$ distribution.
Recall that this distribution, as we obtain it, is
through
calculation and a
choice of
$q_{max}^{soft}$
and $\sigma_0$,
to ensure a good description of the low energy total cross-section.
Hence, our prediction for SPLRG is to
be calculated from the expression
\be
< |S | ^2 > = \int d^2 {\bf b} \ A_{BN}( b ,q_{max}^{soft})
e^{-n_{soft}(b,s)-n_{hard}(b,s)}
\ee

\begin{figure}
  \begin{center}
\includegraphics[scale=0.50]{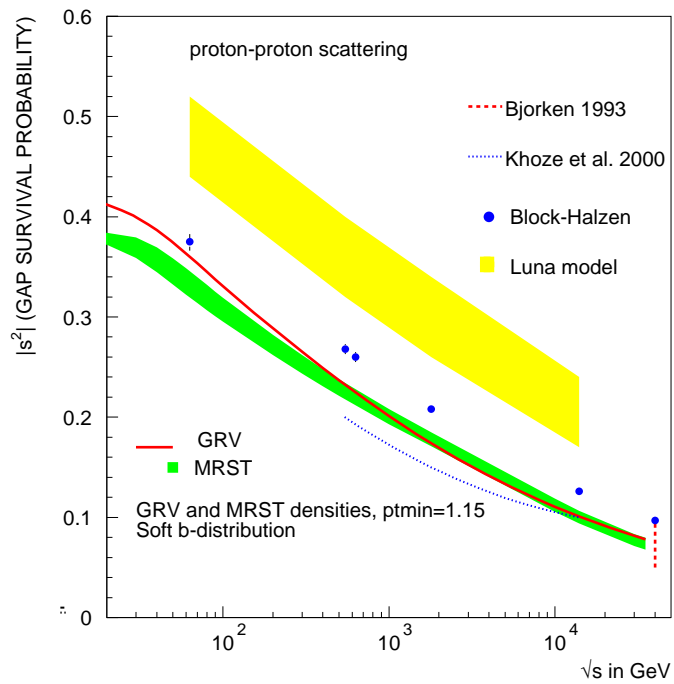}
\caption{Predictions for
survival probability
using different parton densities in
  the GGPS model described in the text and comparison with other models.}
  \label{survband}
\end{center}
\end{figure}

The quantity $q_{max}^{soft}$ has only a very slight energy
dependence, but in principle it can be different for different $p_{tmin}$
and different densities. We calculate  the survival probability for a set
of
parameters and parton densities as used
for the total cross-section. This is given in Figure ~\ref{survband},
where we use MRST and
GRV densities and the set $p_{tmin}=1.15\ GeV$, $\sigma_0=48\ mb$, $p=0.75$.

Our predictions for $<|S^2|>$ are compared here with other similar models,
namely with Luna
~\cite{Luna:2006qp},
with Block and Halzen (BH) ~\cite{Block:2001ru}, with Khoze,
Martin and Ryskin (KMR)
\cite{Khoze:2000cy}
 and with
 Bjorken prediction for SSC energies \cite{Bjorken:1992er}.

Our results for LHC energies  differ from both BH and Luna  models,
however, the difference with BH
is not as pronounced as with the Luna model,
and we would favour a lower value for the survival probability, closer to
the KMR value and within the range predicted by Bjorken. 
Not shown in the figure, there is also the Pythia prediction 
\cite{pythia}, which, in  a multiple scattering formulation 
using CTEQ5 densities within  Rich Field's "Tune A" of Pythia \cite{rich}, 
gives a value 0.026,  which lies lower than all the others.

The comparison with other models shown in Figure 3  indicates a  good
  agreement between different approaches
to the rather loose idea of survival probability.
It is important to notice how all the predictions from QCD
inspired models like BH, BN resummation
like ours, perturbative QCD like KMR
and Bjorken's estimate all fall within a band of  $5 \div 10 \%$.
In our opinion, this convergence of different models lends a strong 
credibility to this type of predictions and puts the concept of Rapidity 
Gaps and studies of their Survival Probability on a rather firm ground. This
increases our confidence in using these to estimate efficacy of existence of 
events with Rapidity Gaps as a means to detect interesting BSM signals 
isolating  them from the background.

\section{Conclusions}
We
have shown above that the range of the results for \sigtot\ from our
GGPS model~\cite{Godbole:2004kx} spans the range of other computations
made using current data and general arguments
based on unitarity and/or factorization. Further, we give
our own estimate of the survival probability for large rapidity gaps at
the LHC and show that our estimates are in reasonable
agreement with other models.
\section*{Acknowledgments}
One of us, A.G., acknowledges partial support  for this work by
MEC (FPA2006-0594) and by Junta de Andaluc\'\i a (FQM 101 and FQM 437).
R.G. would like to thank the theory group at LNF, Frascati for 
hospitality  where part this work was done during a visit.
G.P.  gratefully acknowledges  the hospitality of the 
Theoretical Physics group of
Boston University where part of this work was finished. G.P. is grateful
to J.-R. Cudell, V. Khoze and Leif L\"onnblad for providing numerical 
 estimates for survival probabilities in their models.
Work was also partly
supported by EU contract 2002-00311.

\end{document}